\documentclass[12pt]{article}
\usepackage[utf8]{inputenc} 
\usepackage{amsfonts}
\usepackage{booktabs}
\usepackage{siunitx}

\usepackage{tikz}
\usetikzlibrary{patterns,shapes,decorations}

\usepackage[margin=1in]{geometry}
\usepackage{setspace}
\usepackage{kotex}
\usepackage{graphicx,color} 
\usepackage{amsmath,amssymb}
\usepackage{amsthm}
\usepackage{hyperref} 
\usepackage[export]{adjustbox}
\usepackage{footnote,xcolor}
\hypersetup{linkbordercolor=green}
\usepackage{booktabs} 
\usepackage{array} 
\usepackage{paralist} 
\usepackage{verbatim} 
\usepackage{subfig} 

\usepackage{fancyhdr} 
\usepackage{sectsty}
\allsectionsfont{\sffamily\mdseries\upshape} 

\usepackage[round]{natbib}   
\usepackage{authblk} 

\usepackage[titles,subfigure]{tocloft} 

\newcommand{\E}{\mathbf{E}}

\newcommand{\indep}{\perp \!\!\! \perp}

\newtheorem{assumption}{Assumption}

\newtheorem{theorem}{Theorem}

 \usepackage{titlesec}
 
\usepackage{mdframed}
\usetikzlibrary{shapes,decorations,arrows,calc,arrows.meta,fit,positioning}
\tikzset{
    -Latex,auto,node distance =1 cm and 1 cm,semithick,
    state/.style ={ellipse, draw, minimum width = 0.8 cm},
    point/.style = {circle, draw, inner sep=0.04cm,fill,node contents={}},
    bidirected/.style={Latex-Latex,dashed},
    el/.style = {inner sep=2pt, align=left, sloped}
}

\titleformat*{\section}{\Large\bfseries}
\titleformat*{\subsection}{\large\bfseries}
\titleformat*{\subsubsection}{\large\bfseries}
\titleformat*{\paragraph}{\small\bfseries}
\titleformat*{\subparagraph}{\small}
\date{} 

\usepackage{enumitem}

\usepackage{siunitx}
\usetikzlibrary{positioning}
\tikzset{mynode/.style={draw,text width=1in,align=center}
}

\newlist{thmlist}{enumerate}{1}
\setlist[thmlist]{label=(\roman{thmlisti}),noitemsep}

\usepackage{thmtools}

\usepackage{hyperref}
\hypersetup{colorlinks=true, linkcolor=blue, filecolor=blue, urlcolor=blue, citecolor=blue} 
\usepackage[colorinlistoftodos]{todonotes}

         \renewenvironment{abstract}
 {\small
  \begin{center}
  \bfseries \abstractname\vspace{-.5em}\vspace{0pt}
  \end{center}
  \list{}{%
    \setlength{\leftmargin}{5mm}
    \setlength{\rightmargin}{\leftmargin}%
  }%
  \item\relax}
 {\endlist}

 \newpage

\title{Estimating Spillover Effects \\ in the Presence of Isolated Nodes}

\begin{document}
\author{Bora Kim\thanks{
\noindent Department of Finance, Accounting \& Economics, University of Nottingham Ningbo China, Ningbo 315100, China. \\
E-mail: \texttt{bora.kim@nottingham.edu.cn}.
}}

\date{\today} 
\maketitle

\begin{abstract}
In estimating spillover effects under network interference, practitioners often use linear regression with either the number or fraction of treated neighbors as regressors. An often overlooked fact is that the latter is undefined for units without neighbors (``isolated nodes"). The common practice is to impute this fraction as zero for isolated nodes. This paper shows that such practice introduces bias through theoretical derivations and simulations. Causal interpretations of the commonly used spillover regression coefficients are also provided.
\end{abstract}

\newpage
\section{Introduction}
Consider \(N\) units, indexed by \(i = 1, 2, \ldots, N\), which are connected through a symmetric network \(G\) where \(G_{ij} = 1\) if units \(i\) and \(j\) are direct neighbors, and \(G_{ij} = 0\) otherwise. There are no self-links, so \(G_{ii} = 0\) for all \(i\). Each unit \(i\) is independently assigned a random treatment with probability \(\Pr(D_i=1)=p\in(0,1)\). 

With spatially or network connected data, treatment spillovers chellengages the causal inference When treatment spillovers are present, a unit's outcome \(Y_i\) is influenced not only by its own treatment \(D_i\) but also by the treatments received by others (\cite{cox1958}). This violates a key assumption for causal inference, the Stable Unit Treatment Value Assumption (SUTVA; \cite{rubin1980}).

This influence of others' treatments is often summarized by the number of treated neighbors, \(T_i \equiv \sum_{j \in N_i} D_j\), where \(N_i\) is the set of \(i\)'s direct neighbors, or the proportion of treated neighbors, \(\bar{D}_i \equiv \frac{T_i}{\gamma_i}\), where \(\gamma_i \equiv |N_i|\) denotes the degree of \(i\).

While nonparametric estimators for spillover effects exist, empirical studies typically rely on regression methods.\footnote{See \cite{bryan2014}, \cite{cai2015}, \cite{miguel2004}, \cite{oster2012} for empirical examples using $T$ or $\bar{D}$ as regressors, to name a few. \cite{manski2013}, \cite{leung2020}, \cite{aronow2017} outline theoretical justifications for using such regressors.} Two common regression specifications are:
\begin{align}
\textrm{With}&\quad T_i\equiv \sum_{j \in N_i} D_j,\quad \gamma_i\equiv|N_i|,\quad \bar D_i\equiv \frac{T_i}{\gamma_i}, \notag\\
Y_i &= \alpha_0 + \alpha_d D_i + \alpha_t T_i + \alpha_\gamma \gamma_i + u_i \tag{$T$-regression} \label{T-reg} \\
Y_i &= \eta_0 + \beta_d D_i + \beta_{\bar{d}} \bar{D}_i + u_i \quad \text{using only } \gamma_i > 0 \tag{$\bar{D}$-regression} \label{barD-reg}
\end{align}
The coefficients $\alpha_d$ and $\beta_d$ are meant to capture the ``Direct effect" of one's own treatment, while $\alpha_t$ and $\beta_{\bar{d}}$ capture the ``Spillover effect." However, the exact causal interpretations of these coefficients are unknown. One contribution of this paper is to fill this gap.\footnote{Although \cite{vazquezbare2023} provided causal interpretations for the regression of $Y$ on $(D, \bar{D})$, his framework assumes group interactions and overlooks the network degree. In contrast, our findings emphasize the necessity of adjusting for the degree of the node as a critical observed confounder.}

The first model, referred to as ``$T$-regression," regresses \(Y_i\) on \((D_i, T_i, \gamma_i)\). Controlling for \(\gamma_i\) is essential because, even if \(\{D_i\}\) is random, \(T_i \equiv \sum_{j \in N_i} D_j\) is not, as it depends on \(i\)'s degree, which is endogenous—units with many neighbors would have more treated neighbors, even with random treatment allocation. (\cite{borusyak2023}) Thus, ignoring \(\gamma_i\) can result in omitted variable bias if \(Y_i\) is influenced by \(\gamma_i\). 

In contrast, the second model, referred to as ``$\bar{D}$-regression," regresses \(Y_i\) on \((D_i, \bar{D}_i)\) without controlling for \(\gamma_i\). The rationale is that the proportion of treated neighbors is uncorrelated with a number of one's neighbors, making it unnecessary to control for \(\gamma_i\). We show that this is true \emph{as long as there are no isolated nodes in the sample}. 

An often overlooked complication is that \(\bar{D}_i\) is undefined for units without neighbors (\(\gamma_i = 0\)), commonly referred to as ``isolated nodes." While one might think isolated nodes are a minor issue, they occur frequently in many real-world social networks, which tend to be sparse (\cite{barabasi2016}). For example, both \cite{bandiera2010} and \cite{carter2021} found that about 30\% of their study samples were isolated nodes.

Isolated nodes pose no problem in $T$-regression since we can simply set $T_i = \gamma_i = 0$ for these nodes. However, handling $\bar{D}$-regression becomes tricky when isolated nodes are present. In such cases, researchers typically employ one of two common practices: The first is to exclude isolated nodes and run the $\bar{D}$-regression \emph{using only subsamples of $\gamma_i > 0$} as in \cite{bandiera2010}. We call this ``subsample $\bar{D}$-regression" or simply, ``$\bar D$-regression".

 The second approach is to impute $\bar{D}_i = 0$ for those $\gamma_i = 0$. Let \(\bar{D}_i^*\) denote this imputed value:
\[
\bar{D}_i^* \equiv 
\begin{cases} 
\bar{D}_i \equiv \frac{T_i}{\gamma_i} & \text{if } \gamma_i > 0, \\
0 & \text{if } \gamma_i = 0.
\end{cases}
\]
$Y_i$ is then regressed on $(D_i,\bar D_i^*)$. We call this second approach as $\bar D^*$-regression:
\begin{align}
Y_i &= \eta_0 + \eta_d D_i + \eta_{\bar d} \bar{D}_i^* + u_i \tag{$\bar{D}^*$-regression}
\end{align}
Such an imputation method has been commonly used in literature. For instance, \cite{dupas2014} conducted a randomized price experiment in Kenya, where \(D_i = 1\) if a household received a high subsidy. They used \(\bar{D}^*\)-regression as above, imputing \(\bar{D}_i\) as zero if no study households were within a predefined radius (250 or 500 meters). With a 500-meter radius, about 4\% of the sample were isolated nodes; with a 250-meter radius, about 10\% were isolated nodes. Similarly, \cite{godlonton2012}, who examined peer effects in learning HIV results in rural Malawi, found that approximately 5\% of individuals had no neighbors in the spatial  network. They assigned \(\bar{D}\) (the fraction of neighbors learning their HIV results) as zero, reasoning that \emph{if there are no neighbors, spillover effects should be zero}.

The goal of this paper is to demonstrate that imputing $\bar D$ as zero for isolated nodes introduces bias in the estimation of spillover effects. This is shown in two ways: In Section 2, we illustrate that the $\bar{D}^*$-regression suffers from omitted variable bias, whereas the subsample $\bar{D}$-regression does not. Specifically, when isolated nodes are excluded, it is shown that $\text{Cov}(\bar{D}_i, \gamma_i) = 0$. However, once we use the imputed $\bar{D}_i^*$ for isolated nodes, we have $\text{Cov}(\bar{D}_i^*, \gamma_i) \neq 0$. We argue that this is due to the artificial mass of points created at $\bar{D}_i^* = \gamma_i = 0$.

One might think that including \(\gamma_i\) as an additional covariate in the \(\bar{D}^*\)-regression would resolve the problem. However, we argue that this is incorrect since \emph{spillover effects are undefined, not zero, for isolated nodes}, contrary to practitioners' belief. Thus, the right approach to handling isolated nodes is to exclude them when measuring spillover effects. To demonstrate this, we use the nonparametric potential outcomes framework to rigorously define spillover effects, rather than relying on linear regression specifications, which may not always hold.

We then examine the causal interpretation of OLS coefficients from the $T$-, $\bar{D}$-, and $\bar{D}^*$-regressions.\footnote{Recent discussions, such as by \cite{sloczynski2022}, explore the causal interpretation of OLS coefficients in non-spillover settings. This differs from spillover cases, which involve multiple treatments, including a binary own treatment and a multivalued treatment of neighbors.} We find that the \(T\)- and \(\bar{D}\)-regression coefficients represent valid weighted averages of spillover effects that are heterogeneous in degrees. However, they use different weighting schemes. The \(T\)-regression gives more weight to nodes with higher degrees, with isolated nodes receiving zero weight, which is appropriate as spillover effects are undefined for isolated nodes. In contrast, the subsample \(\bar{D}\)-regression places more weight on nodes with lower degrees, with degree-1 nodes receiving the highest weights.

In contrast, the \(\bar{D}^*\)-regression is biased for any valid weighted averages of spillover effects. This bias is zero only if two conditions hold: the direct treatment effect is not heterogeneous with respect to degree, \emph{and} the degree does not directly affect the outcome. These strong assumptions are unlikely to hold in practice where the network is endogenous (\cite{bramoulle2020}).


Section 4’s simulation study illustrates the extent of this bias involved in the imputation method. We show that $\bar{D}^*$-regression can exhibit sign reversal where the estimated coefficient suggests positive spillover effects even when true spillovers are non-existent or negative for all units. On the other hand, the $T$- and subsample $\bar D$-regressions are shown to be consistent for valid weighted averages of spillover effects.
 
The remainder of this paper is organized as follows: Section 2 demonstrates the omitted variable bias in $\bar{D}^*$-regression. Section 3 uses the potential outcomes framework to show that $\bar{D}^*$-regression coefficient does not identify valid spillover effects. Section 4 presents simulation studies. Section 5 concludes. All proofs are provided in the appendix.

\section{Omitted variable bias in the spillover regression}\label{sec2}
Recall that both $\bar{D}$- and $\bar{D}^*$-regression do not control for the degree $\gamma_i$. The question then arises: \emph{Are $\gamma_i$ correlated with $\bar{D}_i$ or $\bar{D}_i^*$?}  Appendix \ref{ovb} shows that
\[
\text{Cov}(\bar{D}_i^*, \gamma_i) = \left\{\mathbf{E}(\gamma_i | \gamma_i > 0) - \mathbf{E}(\gamma_i)\right\} \mathbf{E}(D_i) \Pr(\gamma_i > 0) > 0.
\]

This covariance is positive because $\mathbf{E}(\gamma_i | \gamma_i > 0) > \mathbf{E}(\gamma_i)$ in the presence of isolated nodes. This implies that the $\bar{D}^*$-regression, which does not control for $\gamma_i$, suffers from omitted variable bias if $Y_i$ is directly affected by $\gamma_i$, which is endogenous. For example, a unit with many neighbors may have a higher outcome regardless of its own and neighbors' treatment status. 

In contrast, when there are no isolated nodes in the data so that \(\bar{D}_i^* = \bar{D}_i\) for all \(i\), the covariance term becomes zero because \(\mathbf{E}(\gamma_i | \gamma_i > 0) = \mathbf{E}(\gamma_i)\) and \(\Pr(\gamma_i = 0)=0\). Therefore, unlike the \(\bar{D}^*\)-regression, the subsample \(\bar{D}\)-regression does not suffer from omitted variable bias, justifying the practice of not controlling for \(\gamma_i\) in the subsample \(\bar{D}\)-regression.

\begin{figure}[h!]
    \centering
    \includegraphics[width=\textwidth]{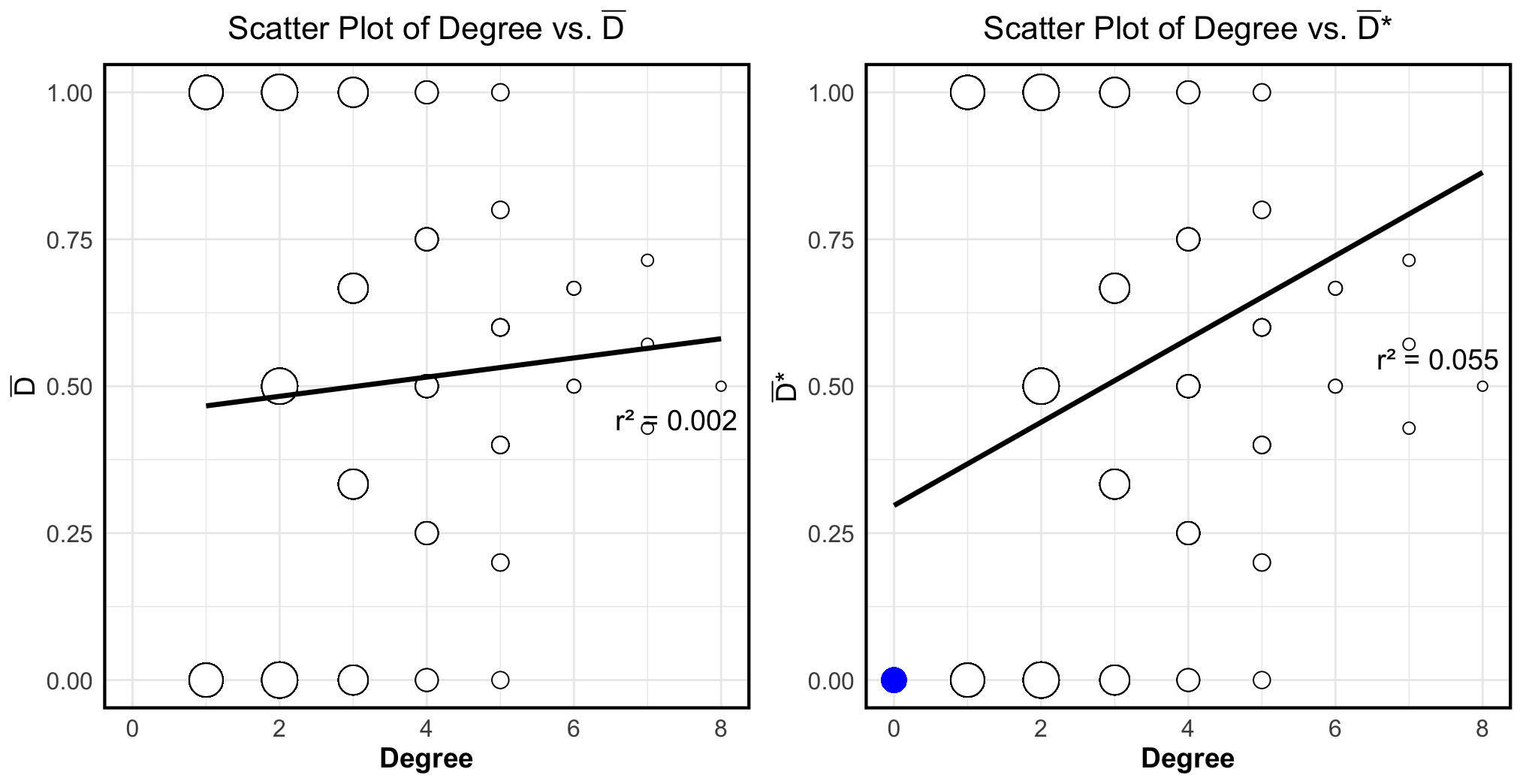}
\caption{Scatterplots of degree vs. $\bar{D}$ (left) and degree vs. imputed $\bar{D}^*$ (right) with regression lines. $r^2$ is the coefficient of determination. Simulated data from a random graph with $N=1,000$ nodes and treatments \(D_i \sim^{\text{iid}} \text{Bernoulli}(0.5)\). Circle size indicates observation frequency. Blue points (right) represent isolated nodes, about 10\% of the observations. }
    \label{fig}
\end{figure}
To illustrate this point, we simulated data with \(N = 1,000\) units using the Watts-Strogatz random graph model, which generates sparse networks with approximately 10\% isolated nodes.\footnote{Watts-Strogatz model is known to generate networks with high clustering and short average path lengths (\cite{aronow2017}).} The treatment assignment was generated \(D_i \sim^{\text{iid}} \text{Bernoulli}(0.5)\), independently of the network generation. The left panel of Figure \ref{fig} shows the scatterplot of \(\gamma_i\) and \(\bar{D}_i\), excluding isolated nodes. The right panel shows the scatterplot of \(\gamma_i\) and \(\bar{D}_i^*\), where isolated nodes ($\gamma_i=\bar D_i^*=0$) are represented as blue circles. 

The first scatterplot of degree vs. \(\bar{D}\) shows no apparent trend, indicated by a nearly flat regression line and a coefficient of determination (\(r^2\)) value close to zero, suggesting that \(\text{Cov}(\gamma_i, \bar{D}_i) = 0\). This is expected since the network and treatment were generated independently. In contrast, the second scatterplot of degree vs. \(\bar{D}^*\) reveals a positive correlation, with an \(r^2\) of 0.055. Given that the only difference between the two datasets used in the scatterplots is the presence of blue circles in the right panel, representing isolated nodes, it is clear that the mass of points created by imputing \(\bar{D}^* = 0\) at \(\gamma = 0\) induces an artificial correlation.
\section{What do spillover coefficients identify?}
In this section, we move from the regression framework to the nonparametric potential outcomes framework to rigorously define spillover effects. We begin by clarifying the implicit assumptions made when researchers use $T$-, $\bar{D}$-, or $\bar{D}^*$-regressions. Based on these assumptions, we define the causal estimands of interest and then examine the causal interpretations of these regression coefficients.

Let \(Y_i(\mathbf{d})\) denote the potential outcome of unit \(i\) under the counterfactual treatment vector \(\mathbf{d}= (d_1, d_2, \cdots, d_N) \in \{0,1\}^N\) (\cite{hudgens2008}). To reduce the dimensionality of \(\bf{d}\), we consider the following assumptions:

\begin{assumption}[Shape Restrictions on Spillovers]\label{a1}\
\begin{enumerate}
    \item[(A1)] Spillovers occur only among direct neighbors, and direct neighbors are exchangeable.
    \item[(A2)] There is no interaction effect between an individual's own treatment and others' treatments on the outcome.
    \item[(A3)] The effect of having one additional treated neighbor is independent of the number of treated neighbors one already has.
\end{enumerate}
\end{assumption}
Assumption (A1) is implicit in conventional regression methods that use the number of treated neighbors or the fraction of treated neighbors as regressors, without considering indirect neighbors. \cite{leung2020} demonstrated that under Assumption (A1), the potential outcomes depend solely on the following sufficient statistics: own treatment ($d$), number of treated neighbors ($t$), and number of neighbors ($\gamma$). Thus, the potential outcome can be written as \(Y_i(d,t,\gamma)\). Since, unlike \((d,t)\), \(\gamma\) is not manipulable, we treat \(\gamma\) as a pre-treatment covariate, and write the potential outcome as \(Y_i(d,t)\) which is defined for \(d \in \{0,1\}\) and \(t \in \{0,1,\cdots,\gamma_i\}\). Here, \(\gamma_i\) is the observed degree of unit \(i\). The observed outcome is \(Y_i = Y_i(D_i, T_i)\).

Causal estimands are defined as follows: For any \(d \in \{0,1\}\) and \(t \in \{0,1,\cdots,\gamma\}\), 
\begin{equation*}
\begin{aligned}
\gamma\text{-conditional Spillover Effect:} \ \lambda^{se}(d,t,\gamma) &\equiv \mathbf{E}\{Y_i(d,t) - Y_i(d,t-1) | \gamma_i = \gamma\}, \\
\gamma\text{-conditional Direct Effect:} \ \mu^{de}(t,\gamma) &\equiv \mathbf{E}\{Y_i(1,t) - Y_i(0,t) | \gamma_i = \gamma\}.
\end{aligned}
\end{equation*}
\(\lambda^{se}(d,t,\gamma)\) quantifies the impact of having an additional treated neighbor while the own treatment is fixed at \(d\) for individuals with the observed degree \(\gamma\). \(\mu^{de}(t,\gamma)\) captures the direct effect of an individual's own treatment for those who have \(t\) treated neighbors out of \(\gamma\) total neighbors.

Assumption (A2) states that \(\lambda^{se}(d,t,\gamma) = \lambda^{se}(t,\gamma)\) and \(\mu^{de}(t,\gamma) = \mu^{de}(\gamma)\). This means the spillover effect of having an additional treated neighbor is independent of an individual's treatment status, and the direct effect of an individual's own treatment is independent of the number of treated neighbors. This assumption is implicit in regression methods that ignore the interaction between \(D\) and \(T\) (or \(\bar{D}\)).

Finally, under Assumption (A3), \(\lambda^{se}(t, \gamma) = \lambda^{se}(\gamma)\). This means that for units with a given degree, the incremental effect of gaining an additional treated neighbor is constant, regardless of the number of existing treated neighbors. This assumption is implied when researchers use \(T_i\) or its normalized version, $\bar D_i$, in regression models, rather than different indicators for each possible value of \(T_i\). \footnote{Relaxing Assumption \ref{a1} would require, for instance, considering linear specifications such as \(Y\) on \((D, T, DT)\), which is beyond the scope of this paper. Note, however, that the proof of Theorem \ref{thm1} in the appendix is first derived assuming only (A1), and then gradually incorporates (A2) and (A3).} 

The randomized treatment allocation assumption is formalized as follows:

\begin{assumption}[Randomized Treatment Allocation]\label{a2}
A binary treatment \(\textbf{D}\equiv\{D_i\}_{i=1}^N\) is allocated randomly across individuals with \(D_i \sim^{\textrm{iid}} \text{Bernoulli}(p)\), where \(p = \mathbf{E}(D_i)\in(0,1)\) with 
$\textbf{D} \indep (\gamma_i, \{Y_i(d,t)\}_{d,t})$ for all $i$.
\end{assumption}
The following theorem provides a ``true" representation of \(Y_i\) under these assumptions:
\begin{theorem} \label{thm1}
Under Assumptions \ref{a1} and \ref{a2}, \(Y_i\) can be represented in a partially linear form as follows:
\begin{equation}\label{eqThm1}
Y_i = \theta^{00}(\gamma_i) + \mu^{de}(\gamma_i)D_i + \lambda^{se}(\gamma_i)T_i + \varepsilon_i, \quad \mathbf{E}(\varepsilon_i | D_i, T_i, \gamma_i) = 0,
\end{equation}
for an unknown \(\gamma\)-conditional intercept \(\theta^{00}(\gamma)\), and \(D\)-, \(T\)-slopes \(\mu^{de}(\gamma), \lambda^{se}(\gamma)\) defined as
\begin{equation*}
\theta^{00}(\gamma_i) \equiv \mathbf{E}\{Y_i(0,0) | \gamma_i\}, \quad \mu^{de}(\gamma_i) \equiv \mathbf{E}\{Y_i(1,t) - Y_i(0,t) | \gamma_i\}, \quad \lambda^{se}(\gamma_i) \equiv \mathbf{E}\{Y_i(d,t) - Y_i(d,t-1) | \gamma_i\}
\end{equation*}
which measure the \(\gamma\)-conditional mean baseline (no intervention) outcome, direct treatment effect, and spillover effect, respectively.
\end{theorem}
Note that equation (\ref{eqThm1}) is derived without assuming any functional form, making it nonparametric. This can be understood as a saturated regression with coefficients being unknown functions of \(\gamma\) (\cite{angrist2009}). In this sense, we refer to this as the ``true" representation.\footnote{A similar nonparametric representation in the non-spillover context is used in \cite{lee2018} and \cite{lee2024}.} This representation suggests that for a subsample of \(\gamma_i = \gamma\), we can apply an OLS of \(Y\) on \((D, T)\) to identify \(\theta^{00}(\gamma)\), \(\mu^{de}(\gamma)\), and \(\lambda^{se}(\gamma)\) nonparametrically. When \(\gamma_i = 0\), this simplifies to \(Y_i = \theta^{00}(0) + \mu^{de}(0)D_i + \varepsilon_i\), indicating that only \(\theta^{00}(0)\) and \(\mu^{de}(0)\) can be identified nonparametrically. This shows that spillovers are not identified for isolated nodes.

While the true representation shows that the effects of \(D_i\) and \(T_i\) on \(Y_i\) are heterogeneous in \(\gamma_i\), the \(T\)-regression assumes that the effect is constant. This raises the question: \emph{What does the OLS coefficient in the regression of \(Y_i\) on \((D_i, T_i, \gamma_i)\) identify, given it assumes constant effects, unlike the representation in equation (\ref{eqThm1})?} The following theorem addresses this.

\begin{theorem}[Interpretation of \(T\)-regression] \label{thm2}
Consider estimating the following linear model $Y_i = \alpha_0 + \alpha_d D_i + \alpha_t T_i + \alpha_{\gamma}\gamma_i + u_i$ using OLS. Under Assumptions \ref{a1} and \ref{a2}, the regression coefficients \(\alpha_d\) and \(\alpha_t\) identify the following:
\begin{equation}
\alpha_d = \mathbf{E}\{\mu^{de}(\gamma_i)\}, \quad \alpha_t = \mathbf{E}\{w_{t}(\gamma_i)\lambda^{se}(\gamma_i)\}, \quad w_{t}(\gamma_i) = \frac{\gamma_i}{\mathbf{E}(\gamma_i)}.
\end{equation}
$w_t(\cdot)\ge 0$ is the weight with $\E\{w_t(\gamma)\}=1$.
\end{theorem}
Thus, the OLS coefficient \(\alpha_d\) identifies the average direct effect, averaged over the distribution of degree \(\gamma\). The \(\alpha_t\) coefficient identifies the weighted average of spillover effects, with the weights being proportional to \(\gamma_i\). Importantly, isolated nodes receive zero weight and therefore do not contribute to \(\alpha_t\). This is appropriate, as spillovers are undefined for isolated nodes.

The following theorem shows that the \(\bar{D}\)-regression, which excludes isolated nodes, also identifies a weighted average of spillover effects but with different weights:
\begin{theorem}[Interpretation of \(\bar D\)-regression]\label{lemma1}
When there are no isolated nodes (or if isolated nodes are excluded), the OLS coefficients from the regression \(Y_i = \beta_0 + \beta_d D_i + \beta_{\bar{d}} \bar{D}_i + u_i\) identify:
\begin{eqnarray}\label{eq3}
\beta_d = \mathbf{E}\{\mu^{de}(\gamma_i)\}, \quad \beta_{\bar{d}} = \mathbf{E}\{w_{\bar{d}}(\gamma_i)\gamma_i\lambda^{se}(\gamma_i)\}, \quad w_{\bar{d}}(\gamma_i) = \frac{1/\gamma_i}{\mathbf{E}(1/\gamma_i)}.
\end{eqnarray}
Here, \(w_{\bar{d}}(\cdot) \ge 0\) is a weight with \(\mathbf{E}\{w_{\bar d}(\gamma)\} = 1\).
\end{theorem}
We see that the $D$-coefficient in the $\bar{D}$-regression identifies the same average direct effect as in the $D$-coefficient of the $T$-regression. The $\bar{D}$-coefficient identifies the weighted averages of $\gamma_i \lambda^{se}(\gamma_i)$, which is the causal effect of a per-unit increase in $\bar{D}_i$ for node with $\gamma_i$. This is because equation (\ref{eqThm1}), when \(\gamma_i > 0\), can be written as:
$Y_i = \theta^{00}(\gamma_i) + \mu^{de}(\gamma_i)D_i + \gamma_i \lambda^{se}(\gamma_i) \bar{D}_i + \varepsilon_i$.  $\gamma_i \lambda^{se}(\gamma_i)$ is then weighted by \(w_{\bar{d}}(\gamma)\), which gives larger weights to nodes with lower degrees, with nodes having a degree of one receiving the largest weight. This inverse weighting scheme is the opposite of that used in Theorem \ref{thm2}. 

The weight \(w_{\bar{d}}(\cdot)\) already implies that \(\gamma_i=0\) cannot be incorporated for $\bar D$-regression. The following theorem demonstrates that, unlike the $T$-, \(\bar{D}\)-regressions, the \(\bar{D}^*\)-regression, which imputes \(\bar{D} = 0\) for isolated nodes, does not identify any valid weighted averages of spillover effects.

\begin{theorem}[Interpretation of \(\bar{D}^*\)-regression]\label{thm3}
Consider estimating the following linear model using OLS: $Y_i = \eta_0 + \eta_d D_i + \eta_{\bar d} \bar{D}_i^* + u_i$ where \(\bar{D}_i^* = \frac{T_i}{\gamma_i}\) if \(\gamma_i > 0\) and \(\bar{D}_i^* = 0\) if \(\gamma_i = 0\). Under Assumptions \ref{a1} and \ref{a2}, the regression coefficients \(\eta_d\) and \(\eta_{\bar d}\) identify:
\begin{align}\label{eq4}
\eta_d = \mathbf{E}\{\mu^{de}(\gamma_i)\}, \quad
\eta_{\bar d} = \text{bias} + \frac{\mathbf{E}\big\{\gamma_i \lambda^{se}(\gamma_i)\bar{D}_i [\bar{D}_i - \mathbf{E}(\bar{D}_i^*)]\mid \gamma_i > 0\big\}}{\mathbf{E}\big\{\bar{D}_i [\bar{D}_i - \mathbf{E}(\bar{D}_i^*)] \mid \gamma_i > 0\big\}}
\end{align}
where
\begin{eqnarray*}
\text{bias} &=& \frac{(\Delta \theta^{00} + p \Delta \mu^{de})(1 - p_{\gamma})}{p(1 - p_{\gamma}) + (1 - p)\mathbf{E}\left(\frac{1}{\gamma_i} \mid \gamma_i > 0\right)}; \ p \equiv \Pr(D_i=1), \quad p_{\gamma} \equiv \Pr(\gamma_i > 0), \\
\Delta \theta^{00} &\equiv& \mathbf{E}\{\theta^{00}(\gamma_i) \mid \gamma_i > 0\} - \mathbf{E}\{\theta^{00}(\gamma_i) \mid \gamma_i = 0\}, \\
\Delta \mu^{de} &\equiv& \mathbf{E}\{\mu^{de}(\gamma_i) \mid \gamma_i > 0\} - \mathbf{E}\{\mu^{de}(\gamma_i) \mid \gamma_i = 0\}.
\end{eqnarray*}

\end{theorem}

Theorem~\ref{thm3} establishes that while \(\eta_d\) identifies the same average direct effect as in $T$- and $\bar D$-regression, \(\eta_{\bar d}\) consists of two parts: a bias component and a component that involves the weighted average of the spillover effect, captured by \(\gamma_i \lambda^{se}(\gamma_i)\). While the second term can be shown to have a positive weight that with mean of 1, it is hard to interpret this weight due to \(\bar{D}_i\) being incorrectly centered by \(\mathbf{E}(\bar{D}_i^*) = p \cdot p_{\gamma}\), differently from \(\mathbf{E}(\bar{D}_i) = p\).

Importantly, \(\eta_{\bar d}\) is not a pure measure of spillover effects but is contaminated by a bias term. This bias arises from differences in the baseline outcome (\(\Delta \theta^{00}\)) and the direct effect (\(\Delta \mu^{de}\)) between isolated and non-isolated units. The bias is zero only if both \(\Delta \theta^{00} = 0\) \emph{and} \(\Delta \mu^{de} = 0\), indicating no heterogeneity in the baseline outcome $Y_i(0,0)$, and in the direct effect across these units. The former implies no direct impact of degree on the outcome, as discussed in Section 2 regarding omitted variable bias, while the latter suggests the direct treatment effect is constant regardless of degree. 

Such a lack of effect heterogeneity is clearly restrictive. For example, units with higher degrees might have greater outcomes even without intervention (\(d=t=0\)) if popularity is correlated with \(Y\). Additionally, they may experience larger direct treatment effects, possibly due to better understanding the benefits of treatment from their neighbors. These scenarios violate the assumption of no effect heterogeneity.

The direction of the bias depends on the signs of \(\Delta \theta^{00}\) and \(\Delta \mu^{de}\), making it difficult to determine a priori whether the bias will be positive or negative. Note that as the proportion of non-isolated nodes, \(p_{\gamma}\), increases, the magnitude of the absolute bias diminishes. When \(p_{\gamma} = 1\) (i.e., no isolated nodes), the bias is zero, and the \(\bar{D}^*\)-regression becomes equivalent to the \(\bar{D}\)-regression, where \(\eta_{\bar{d}}\) in equation \ref{eq4} becomes \(\beta_{\bar{d}}\) in equation \ref{eq3} as shown in the proof.



\section{Simulation demonstrations}
The goal of this section is twofold: First, we demonstrate the bias of $\bar{D}^*$-regression by showing that even when the true spillover effect is zero or negative for all units, the imputation method can yield significantly positive spillover estimates. Second, we verify that our derivations in Theorem~\ref{thm2}, \ref{lemma1},  \ref{thm3} are correct.

\paragraph{Data Generating Process} The total number of simulation repetitions is 5,000. For each simulation, we generate a random graph with \(N=1,000\) units from the Watts-Strogatz  model as in Section 2. Approximately 10\% of the nodes are isolated. The mean degree is 2, and the maximum degree averaging around 7. The treatment assignment \(D_i\) is generated as \(D_i \sim^{\text{iid}} \text{Bernoulli}(0.5)\), and the error term \(\varepsilon_i\) is generated as \(\varepsilon_i \sim^\text{iid} N(0,1)\) with \(\varepsilon_i \indep D_i\) for all \(i\). We consider three designs for generating \(Y_i\):
\begin{align*}
\text{Design 1:} & \quad Y_i = 1 + \gamma_i + D_i + \frac{c}{1 + \gamma_i}T_i + \varepsilon_i, \\
\text{Design 2:} & \quad Y_i = 1 + 1\{\gamma_i > 0\} + \frac{c}{1 + \gamma_i}T_i + \varepsilon_i, \\
\text{Design 3:} & \quad Y_i = 1 + D_i + \frac{c}{1 + \gamma_i}T_i + \varepsilon_i, \\
& \quad \text{where } c = 0 \text{ (zero spillover)} \text{ or} -0.5 \text{ (negative spillover)}.
\end{align*}

Recalling that \(\Delta \theta^{00} \equiv \mathbf{E}\{Y_i(0,0) | \gamma_i > 0\} - \mathbf{E}\{Y_i(0,0) | \gamma_i = 0\}\) controls the bias of $\bar D^*$-regression, Designs 1-3 yield \(\Delta \theta^{00} \approx 2.24\), \(\Delta \theta^{00} = 1\), and \(\Delta \theta^{00} = 0\), respectively. Aside from \(\Delta \theta^{00}\), all three designs have the same true direct effect and spillover effects.

The individual direct effect \(Y_i(1,t) - Y_i(0,t)\) is set to 1 for all units, making the true direct effect \(\mu^{de}(\gamma) \equiv \mathbf{E}\{Y_i(1,t) - Y_i(0,t) | \gamma_i = \gamma\} = 1\) and \(\Delta \mu^{de} \equiv \mathbf{E}\{\mu^{de}(\gamma_i) | \gamma_i > 0\} - \mathbf{E}\{\mu^{de}(\gamma_i) | \gamma_i = 0\} = 0\) in all designs.

The true spillover effects are \(\lambda^{se}(\gamma) \equiv \mathbf{E}\{Y_i(0,t) - Y_i(0,t-1) | \gamma_i = \gamma\} = \frac{c}{1+\gamma}\), where \(c\) is either 0 or -0.5. For \(c = 0\), spillover effects are zero for all units. When \(c = -0.5\), spillover effects are negative for all units and are heterogeneous in \(\gamma\), ranging from -0.5 (\(\gamma = 0\)) to -0.0625 (\(\gamma = 7\)), with the magnitude of spillover effects decreasing (in absolute terms) as nodes have higher degrees.

\paragraph{Estimators} We evaluate three specifications:
\begin{align}
Y_i &= \alpha_0 + \alpha_d D_i + \alpha_t T_i + \alpha_\gamma \gamma_i + u_i \tag{$T$-regression} \\
Y_i &= \beta_0 + \beta_d D_i + \beta_{\bar d} \bar{D}_i + u_i, \quad \text{using only } \gamma_i > 0 \tag{$\bar{D}$-regression} \\
Y_i &= \eta_0 + \eta_d D_i + \eta_{\bar d} \bar{D}_i^* + u_i \tag{$\bar{D}^*$-regression}
\end{align}
We numerically compute the population OLS coefficients \((\alpha_d, \alpha_t, \beta_d, \beta_{\bar{d}}, \eta_d, \eta_{\bar{d}})\) based on Theorem \ref{thm2}, \ref{lemma1}, \ref{thm3} in each case. These are given in the ``True coefficient" part of Table \ref{tab:results}. A caveat is needed for the spillover coefficient \(\eta_{\bar{d}}\) of the \(\bar{D}^*\)-regression. Theorem \ref{thm3} suggests that \(\eta_{\bar{d}}\) is composed of two parts: the bias part and the weighted average part. We take the weighted average part as the ``true coefficient" in the sense that this is the target parameter that \(\eta_d\) is intended to measure.

For example, in Table \ref{tab:results}, when \(c=0\) so that spillover effects are zero for all units, \(\alpha_t = \beta_{\bar{d}} = 0\) as they measure the weighted averages of spillovers, which are zero for all units. On the other hand, the bias part of \(\eta_{\bar d}\) is approximately 0.704, while the weighted average part of \(\eta_{\bar{d}}\) is zero. Hence, the ``true coefficient" for \(\eta_{\bar{d}}\) is reported as 0 in Table \ref{tab:results}. For direct effect coefficients, in all cases, the theorems predict that \(\alpha_d = \beta_d = \eta_d = 1\) since the direct effects are 1 for all units.

The OLS estimates of each regression, averaged across 5,000 simulations, are reported under ``Estimated coefficient.'' Bias is computed as the estimated coefficient minus the true coefficient. The 95\% confidence interval (CI) is calculated as the estimated coefficient $\pm 1.96 \times \text{standard error}$. We use the standard error assuming iid data, which is appropriate for our data-generating process.

 \paragraph{Results} When \(c = 0\) and the true spillover effect is zero, both the $T$ and $\bar{D}$ regressions yield spillover coefficients close to zero, with negligible bias across all designs. The 95\% confidence intervals correctly include the true value of zero. However, the $\bar{D}^*$ regression, which imputes zero for isolated nodes, exhibits significant bias. Design 1, with the highest $\Delta \theta^{00}$, shows the largest bias, with an average estimated coefficient around 0.703 and 95\% confidence intervals that never include the true value of zero. Design 2, with a medium $\Delta \theta^{00}$, also shows bias of 0.314, though the bias is smaller than in Design 1 as expected. Only in Design 3, where \(\Delta \theta^{00} = 0\), are the spillover coefficients unbiased.
 
When \(c = -0.5\), the true spillover effects are heterogeneous in degree but negative for all units. Each regression model combines these heterogeneous effects differently. Numerical computation of the ``true coefficient" yields \(\alpha_t = -0.146\) and \(\beta_{\bar{d}} = -0.298\). For \(\eta_{\bar{d}}\), which includes a bias plus a weighted average part, we take the latter as the ``true coefficient," which is \(\eta_{\bar{d}} = -0.303\), similar to \(\beta_{\bar{d}}\). The table shows that the $T$ and $\bar{D}$ regressions are unbiased for these parameters across all designs. In contrast, the \(\bar{D}^*\) regression again produces biased spillover effects. In Design 1, it shows positive spillover estimates; Design 2 is less biased but still shows positive estimates, with 95\% confidence intervals not including the true coefficient. In Design 3, there is no bias.

For direct effects, all regressions across all designs deliver unbiased estimates for the true direct effects of 1, indicating that the direct effect estimates are unaffected by the regression specifications, as predicted.

\newpage
\begin{table}[htbp]
    \centering
    \caption{Simulation Results for $c=0$ (zero spillovers) and $c=-0.5$ (negative spillovers)}
    \resizebox{\textwidth}{!}{%
    \begin{tabular}{llcccccc}
        \toprule
        & & \multicolumn{3}{c}{Spillover Coefficients} & \multicolumn{3}{c}{Direct Effect Coefficients} \\
        \cmidrule(lr){3-5} \cmidrule(lr){6-8}
        & & $T$-reg & $\bar{D}$-reg & $\bar{D}^*$-reg & $T$-reg & $\bar{D}$-reg & $\bar{D}^*$-reg \\
        \midrule
        \multicolumn{8}{c}{\textbf{$c=0$}: Case of Zero Spillovers} \\
        \midrule
        \multicolumn{8}{l}{\textbf{Design 1: $Y_i = 1 + \gamma_i + D_i + \varepsilon_i$ with $\Delta \theta^{00} = 2.24$}} \\
        \midrule
        & Estimated Coefficient & 0.000 & -0.003 & 0.703 & 1.001 & 1.002 & 1.001 \\
        & True Coefficient & 0 & 0 & 0 & 1 & 1 & 1 \\
        & Bias & 0.000 & -0.003 & 0.703 & 0.001 & 0.002 & 0.001 \\
        & 95\%-CI & (-0.088, 0.088) & (-0.267, 0.261) & (0.448, 0.959) & (0.877, 1.125) & (0.801, 1.203) & (0.801, 1.202) \\
        \midrule
        \multicolumn{8}{l}{\textbf{Design 2: $Y_i = 1 + 1\{\gamma_i > 0\} + \varepsilon_i$ with $\Delta \theta^{00} = 1$}} \\
        \midrule
        & Estimated Coefficient & 0.000 & 0.000 & 0.314 & 1.000 & 1.001 & 1.001 \\
        & True Coefficient & 0 & 0 & 0 & 1 & 1 & 1 \\
        & Bias & 0.000 & 0.000 & 0.314 & 0.000 & 0.001 & 0.001 \\
        & 95\%-CI & (-0.091, 0.091) & (-0.173, 0.173) & (0.150, 0.479) & (0.872, 1.129) & (0.869, 1.132) & (0.872, 1.130) \\
        \midrule
        \multicolumn{8}{l}{\textbf{Design 3: $Y_i = 1 + D_i + \varepsilon_i$ with $\Delta \theta^{00} = 0$}} \\
        \midrule
        & Estimated Coefficient & 0.000 & 0.000 & -0.000 & 1.001 & 1.001 & 1.001 \\
        & True Coefficient & 0 & 0 & 0 & 1 & 1 & 1 \\
        & Bias & 0.000 & 0.000 & -0.000 & 0.001 & 0.001 & 0.001 \\
        & 95\%-CI & (-0.088, 0.088) & (-0.173, 0.173) & (-0.159, 0.158) & (0.877, 1.125) & (0.869, 1.132) & (0.877, 1.125) \\
        \midrule
        \multicolumn{8}{c}{\textbf{$c=-0.5$}: Case of Negative Spillovers} \\
        \midrule
        \multicolumn{8}{l}{\textbf{Design 1: $Y_i = 1 + \gamma_i + D_i - \frac{0.5}{1 + \gamma_i} T_i + \varepsilon_i$ with $\Delta \theta^{00} = 2.24$}} \\
        \midrule
        & Estimated Coefficient & -0.146 & -0.301 & 0.401 & 1.001 & 1.002 & 1.001 \\
        & True Coefficient & -0.146 & -0.298 & -0.303 & 1 & 1 & 1 \\
        & Bias & 0.000 & -0.003 & 0.703 & 0.001 & 0.002 & 0.002 \\
        & 95\%-CI & (-0.234, -0.058) & (-0.562, -0.041) & (0.148, 0.653) & (0.876, 1.125) & (0.804, 1.200) & (0.804, 1.199) \\
        \midrule
        \multicolumn{8}{l}{\textbf{Design 2: $Y_i = 1 + 1\{\gamma_i > 0\} + D_i - \frac{0.5}{1 + \gamma_i} T_i + \varepsilon_i$ with $\Delta \theta^{00} = 1$}} \\
        \midrule
        & Estimated Coefficient & -0.146 & -0.298 & 0.012 & 1.000 & 1.001 & 1.001 \\
        & True Coefficient & -0.146 & -0.298 & -0.303 & 1 & 1 & 1 \\
        & Bias & 0.000 & 0.000 & 0.314 & 0.000 & 0.001 & 0.001 \\
        & 95\%-CI & (-0.236, -0.055) & (-0.471, -0.125) & (-0.153, 0.177) & (0.873, 1.128) & (0.869, 1.132) & (0.872, 1.130) \\
        \midrule
        \multicolumn{8}{l}{\textbf{Design 3: $Y_i = 1 + D_i - \frac{0.5}{1 + \gamma_i} T_i + \varepsilon_i$ with $\Delta \theta^{00} = 0$}} \\
        \midrule
        & Estimated Coefficient & -0.146 & -0.298 & -0.303 & 1.001 & 1.001 & 1.001 \\
        & True Coefficient & -0.146 & -0.298 & -0.303 & 1 & 1 & 1 \\
        & Bias & 0.000 & 0.000 & -0.000 & 0.001 & 0.001 & 0.001 \\
        & 95\%-CI & (-0.234, -0.058) & (-0.471, -0.125) & (-0.461, -0.144) & (0.876, 1.125) & (0.869, 1.132) & (0.876, 1.125) \\
        \bottomrule
    \end{tabular}%
    }
    \caption*{
     5,000 simulations with $N=1,000$. Bias = Estimated coefficient - True coefficient. 95\%-CI is [Estimated coefficient $\pm$ 1.96se]. \\
    \textbf{$T$-reg:} $Y = \beta_0 + \beta_1 D + \beta_2 T + \beta_3 \gamma + u$.\\
    \textbf{$\bar{D}$-reg:} $Y = \beta_0 + \beta_1 D + \beta_2 \bar{D} + u$ after excluding isolated nodes.\\
    \textbf{$\bar{D}^*$-reg:} $Y = \beta_0 + \beta_1 D + \beta_2 \bar{D}^* + u$ with $\bar{D}^*$ imputed as zero for isolated nodes.\\
    Spillover coefficients are $\beta_2$, direct effect coefficients are $\beta_1$.
}
    \label{tab:results}
\end{table}

\newpage
\section{Concluding remarks}

Network-based randomized experiments are gaining popularity for their ability to estimate the spillover effects of neighbors' treatment on an individual's outcome. While theoretical papers mostly focus on nonparametric estimators, many applied studies rely on linear models. The commonly used linear specifications include (1) the $T$-regression, which regresses $Y$ on $D$ (own treatment) and $T$ (the number of treated neighbors) after controlling for the network size $\gamma$ (the number of neighbors), and (2) the $\bar{D}$-regression, which regresses $Y$ on $D$ and $\bar{D}$ (the proportion of treated neighbors). Although coefficients on $T$ and $\bar{D}$ are expected to measure certain spillover effects, rigorous derivations of the spillover effects identified by these linear specifications, as well as the assumptions required therein, are unexplored in the literature.

In this paper, based on the potential outcomes framework, we derived a nonparametric representation of the outcome as a partially linear function of \(D\) and \(T\), each with coefficients as unknown functions of degree \(\gamma\). It is then shown that the $T$-coefficient in (1) identifies a weighted average of spillover effects, where nodes with higher degrees are given larger weights, whereas the $\bar{D}$-coefficient in (2) identifies a weighted average of spillover effects, where nodes with smaller degrees are given more weights. 

Furthermore, the commonly used imputation method in (2), where \(\bar{D}\) is imputed as zero for isolated nodes, introduces significant bias in spillover effect estimates. This bias is zero only under two restrictive conditions: no direct effect of degree on the outcome, and no degree-heterogeneous direct effect. Our simulation studies demonstrate that such imputation can falsely suggest significantly positive spillover effects when there are none, or even negative spillover effects for all units.  Although isolated nodes frequently appear in practice, there has been no rigorous guidance on how to properly handle them. This paper fills this gap.


This study did not address issues of network sampling and network misspecification. In practice, when some units report having zero friends, this may not necessarily mean they truly have no friends, but rather reflect mismeasurement of peer variables. While the topic of mismeasured network data and its implications for linear models is crucial, as demonstrated by \cite{griffith2021}, it is beyond the scope of this study.

Additionally, this paper imposed several shape restrictions on spillovers, such as exchangeable neighbors and the absence of interaction effects between own treatment and others' treatments. While these assumptions are widely used in practice, they can be violated. Given the current interest in the consequences of misspecified exposures in the literature (\cite{savje2024}), this could be an interesting direction for future work.

\newpage
\bibliographystyle{plainnat}

\begin{thebibliography}{}

\bibitem[Angrist \& Pischke, 2009]{angrist2009}
Angrist, J.D., \& Pischke, J.S. (2009). \textit{Mostly harmless econometrics}. Princeton University Press.

\bibitem[Aronow \& Samii, 2017]{aronow2017}
Aronow, P. M., \& Samii, C. (2017). Estimating average causal effects under general interference. \textit{Annals of Applied Statistics}, 11(4), 1912-1947.

\bibitem[Aronow et al., 2021]{aronow2021}
Aronow, P. M., Eckles, D., Samii, C., \& Zonszein, S. (2021). Spillover Effects in Experimental Data. In D. P. Green \& J. N. Druckman (Eds.), \textit{Advances in Experimental Political Science}. Cambridge: Cambridge University Press.

\bibitem[Bandiera et al., 2010]{bandiera2010}
Bandiera, O., Barankay, I., \& Rasul, I. (2010). Social incentives in the workplace. \textit{The Review of Economic Studies}, 77, 417–458.

\bibitem[Barabási, 2016]{barabasi2016}
Barabási, A.-L. (2016). \textit{Network Science}. Cambridge: Cambridge University Press. ISBN: 9781107076266, 1107076269. 

\bibitem[Borusyak \& Hull, 2023]{borusyak2023}
Borusyak, K., \& Hull, P. (2023). Nonrandom Exposure to Exogenous Shocks. \textit{Econometrica}, 91: 2155-2185.

\bibitem[Bramoullé et al., 2020]{bramoulle2020}
Bramoullé, Y., Djebbari, H., \& Fortin, B. (2020). Peer Effects in Networks: A Survey. \textit{Annual Review of Economics}, 12, 603-629. 


\bibitem[Bryan et al., 2014]{bryan2014}
Bryan, G., Chowdhury, S., and Mobarak, A. M. (2014). Underinvestment in a Profitable Technology: The Case of Seasonal Migration in Bangladesh. \textit{Econometrica}, 82(5), 1671-1748.

\bibitem[Cai et al., 2015]{cai2015}
Cai, J., De Janvry, A., and Sadoulet, E. (2015). Social Networks and the Decision to Insure. \textit{American Economic Journal: Applied Economics}, 7(2), 81–108.

\bibitem[Carter et al., 2021]{carter2021}
Carter, M., Laajaj, R., \& Yang, D. (2021). Subsidies and the African Green Revolution: Direct Effects and Social Network Spillovers of Randomized Input Subsidies in Mozambique. \textit{American Economic Journal: Applied Economics}, 13(2), 206-29.

\bibitem[Cox, 1958]{cox1958}
Cox, D. R. 1958. Planning of Experiments. Wiley.

\bibitem[Dupas, 2014]{dupas2014}
Dupas, P. (2014). Short-Run Subsidies and Long-Run Adoption of New Health Products: Evidence From a Field Experiment. \textit{Econometrica}, 82, 197–228.

\bibitem[Godlonton \& Thornton, 2012]{godlonton2012}
Godlonton, S., \& Thornton, R. (2012). Peer effects in learning HIV results. \textit{Journal of Development Economics}, 97, 118–129.

\bibitem[Griffith, 2021]{griffith2021}
Griffith, A. (2021). Name your friends, but only five? The importance of censoring in peer effects estimates using social network data. \textit{Journal of Labor Economics}, 40(4), 779–805.

\bibitem[Hudgens \& Halloran, 2008]{hudgens2008}
Hudgens, M. G., \& Halloran, M. E. (2008). Toward causal inference with interference. \textit{Journal of the American Statistical Association}, 103(482), 832-842.

\bibitem[Lee, 2018]{lee2018}
Lee, M.J. (2018). Simple least squares estimator for treatment effects using propensity score residuals. \textit{Biometrika}, 105, 149-164.

\bibitem[Lee \& Han, 2024]{lee2024}
Lee, M.J., \& Han, C. (2024). Ordinary least squares and instrumental variable estimators for any outcome and heterogeneity. \textit{Stata Journal}, forthcoming.

\bibitem[Leung, 2020]{leung2020}
Leung, M. P. (2020). Treatment and Spillover Effects Under Network Interference. \textit{The Review of Economics and Statistics}, 102(2), 368–380.


\bibitem[Manski, 2013]{manski2013}
Manski, C. F. (2013). Identification of Treatment Response with Social Interactions. \textit{The Econometrics Journal}, 16, S1-S23.

\bibitem[Miguel \& Kremer, 2004]{miguel2004}
Miguel, E., \& Kremer, M. (2004). Worms: Identifying Impacts on Education and Health in the Presence of Treatment Externalities. \textit{Econometrica}, 72, 159–217.

\bibitem[Oster \& Thornton, 2012]{oster2012}
Oster, E., \& Thornton, R. (2012). Determinants Of Technology Adoption: Peer Effects In Menstrual Cup Take-Up. \textit{Journal of the European Economic Association}, 10, 1263–1293.

\bibitem[Rubin, 1980]{rubin1980}
Rubin, D. B. (1980). Comment on ”Randomization Analysis of Experimental Data in the Fisher Randomization Test” by D. Basu. \textit{Journal of the American Statistical Association}, 75, 591–593.

\bibitem[Sävje, 2024]{savje2024}
Sävje, F. (2024). Causal inference with misspecified exposure mappings: separating definitions and assumptions. \textit{Biometrika}, 111(1), 1–15.

\bibitem[Słoczyński, 2022]{sloczynski2022}
Słoczyński, T. (2022). Interpreting OLS Estimands When Treatment Effects Are Heterogeneous: Smaller Groups Get Larger Weights. \textit{The Review of Economics and Statistics}, 104(3), 501–509.

\bibitem[Vazquez-Bare, 2023]{vazquezbare2023}
Vazquez-Bare, G. (2023). Identification and estimation of spillover effects in randomized experiments. \textit{Journal of Econometrics}, 237(1), 105237. ISSN 0304-4076.

\end{thebibliography}


\newpage
\begin{appendix}
\section*{Appendix}
\section{Proofs}

\subsection{Proof for Omitted Variable Bias in Section 2}\label{ovb}
Note that
\[
\text{Cov}(\bar{D}_i^*, \gamma_i) = \mathbf{E}(\bar{D}_i^* \gamma_i) - \mathbf{E}(\bar{D}_i^*) \mathbf{E}(\gamma_i).
\]
Consider the first term on the right-hand side:
\begin{align*}
\mathbf{E}(\bar{D}_i^* \gamma_i) &= \mathbf{E}(\bar{D}_i^* \gamma_i \mid \gamma_i > 0) \Pr(\gamma_i > 0) \\
&= \mathbf{E}(T_i \mid \gamma_i > 0) \Pr(\gamma_i > 0) \\
&= \mathbf{E}\left[\mathbf{E}(T_i \mid \gamma_i > 0, \gamma_i) \mid \gamma_i > 0\right] \Pr(\gamma_i > 0) \\
&= \mathbf{E}(\gamma_i \mid \gamma_i > 0) \mathbf{E}(D_i) \Pr(\gamma_i > 0),
\end{align*}
where the first and third equality use the law of total probability, and the last equality follows from \(\mathbf{E}(T_i \mid \gamma_i > 0, \gamma_i) = \mathbf{E}(T_i \mid \gamma_i) = \gamma_i \mathbf{E}(D_i).\)

While,
\[
\mathbf{E}(\bar{D}_i^*) \mathbf{E}(\gamma_i) = \mathbf{E}(D_i) \mathbf{E}(\gamma_i) \Pr(\gamma_i > 0).
\]
Since \(\mathbf{E}(\bar{D}_i^*) = \mathbf{E}(\bar{D}_i \mid \gamma_i > 0) \Pr(\gamma_i > 0)\) by the law of total probability, and that \(\bar{D}_i^* = \bar{D}_i\) for \(\gamma_i > 0\). Also \(\mathbf{E}(\bar{D}_i \mid \gamma_i > 0) = \mathbf{E}\left\{\mathbf{E}(\bar{D}_i \mid \gamma_i, \gamma_i > 0) \mid \gamma_i > 0\right\}\) whereas \(\mathbf{E}(\bar{D}_i \mid \gamma_i, \gamma_i > 0) = \mathbf{E}(D_i)\).

Combining these,
\[
\text{Cov}(\bar{D}_i^*, \gamma_i) = \left\{\mathbf{E}(\gamma_i \mid \gamma_i > 0) - \mathbf{E}(\gamma_i)\right\} \mathbf{E}(D_i) \Pr(\gamma_i > 0) 
\]
as claimed.

\subsection{Proof for Theorem~\ref{thm1}} \label{proof:thm1}

We first outline the general representation of \(Y_i\) under Assumptions \ref{a1}(A1) and \ref{a1}(A2). We then present the outcome representation further assuming (A3). Assumption \ref{a2} (randomized treatment) is maintained throughout.

\subsection*{General Representation under Assumption \ref{a1}(A1) and \ref{a2}}
Under Assumption \ref{a1}(A1), \cite{leung2020} shows the potential outcome can be written as \(Y_i(d, t)\), where \(t\) is the number of treated neighbors, defined for \(t = 0, 1, 2, \ldots, \gamma_i\) . The observed outcome is:
\begin{equation}\label{eqY}
Y_i = Y_i(D_i, T_i) = D_i Y_i(1, T_i) + (1 - D_i) Y_i(0, T_i) = Y_i(0, T_i) + D_i \left( Y_i(1, T_i) - Y_i(0, T_i) \right).
\end{equation}
First, \(Y_i(0, T_i)\) can be rewritten as follows:
\[
Y_i(0, T_i) = Y_i(0, 0) + \sum_{t = 1}^{\gamma_i} 1(T_i \ge t) \left( Y_i(0, t) - Y_i(0, t - 1) \right).
\]
Similarly for \(Y_i(1, T_i)\):
\[
Y_i(1, T_i) = Y_i(1, 0) + \sum_{t = 1}^{\gamma_i} 1(T_i \ge t) \left( Y_i(1, t) - Y_i(1, t - 1) \right).
\]
Thus,
\[
Y_i(1, T_i) - Y_i(0, T_i) = Y_i(1, 0) - Y_i(0, 0) + \sum_{t = 1}^{\gamma_i} 1(T_i \ge t) \left[ \left( Y_i(1, t) - Y_i(1, t - 1) \right) - \left( Y_i(0, t) - Y_i(0, t - 1) \right) \right].
\]
Substituting these, equation \eqref{eqY} becomes
\begin{align*}
Y_i &= Y_i(0, 0) + \sum_{t = 1}^{\gamma_i} 1(T_i \ge t) \left( Y_i(0, t) - Y_i(0, t - 1) \right) \\
&\quad + D_i \left\{ Y_i(1, 0) - Y_i(0, 0) + \sum_{t = 1}^{\gamma_i} 1(T_i \ge t) \left[ \left( Y_i(1, t) - Y_i(1, t - 1) \right) - \left( Y_i(0, t) - Y_i(0, t - 1) \right) \right] \right\}.
\end{align*}

Taking \(\mathbf{E}(\cdot \mid D_i, T_i, \gamma_i)\) and noting that \(\mathbf{E}(Y_i(d, t) \mid D_i, T_i, \gamma_i) = \mathbf{E}(Y_i(d, t) \mid \gamma_i)\) under Assumption \ref{a2}, we have:
\begin{align}
\mathbf{E}(Y_i \mid D_i, T_i, \gamma_i) &= \theta^{00}(\gamma_i) + \sum_{t = 1}^{\gamma_i} 1(T_i \ge t) \lambda^{se}_0(t, \gamma_i) \notag \\
&\quad + D_i \left\{ \mu^{de}(\gamma_i) + \sum_{t = 1}^{\gamma_i} 1(T_i \ge t) \left( \lambda^{se}_1(t, \gamma_i) - \lambda^{se}_0(t, \gamma_i) \right) \right\} \label{condE}.
\end{align}
where
\begin{eqnarray*}
\theta^{00}(\gamma_i) &\equiv& \mathbf{E}[Y_i(0, 0) \mid \gamma_i], \\
\mu^{de}(\gamma_i) &\equiv& \mathbf{E}[Y_i(1, 0) - Y_i(0, 0) \mid \gamma_i], \\
\lambda^{se}_0(t, \gamma_i) &\equiv& \mathbf{E}[Y_i(0, t) - Y_i(0, t - 1) \mid \gamma_i], \\
\lambda^{se}_1(t, \gamma_i) &\equiv& \mathbf{E}[Y_i(1, t) - Y_i(1, t - 1) \mid \gamma_i].
\end{eqnarray*}

\subsection*{Representation under Additional Assumptions \ref{a1}-(A2) and (A3)}
Under Assumption \ref{a1}-(A2), which posits no interaction between \(D_i\) and \(T_i\), it follows that \(\lambda^{se}_0(t, \gamma_i) = \lambda^{se}_1(t, \gamma_i)\). We denote this common function by \(\lambda^{se}(t, \gamma_i)\). Consequently, equation \eqref{condE} simplifies to:
\begin{equation}\label{condE2}
\mathbf{E}(Y_i \mid D_i, T_i, \gamma_i) = \theta^{00}(\gamma_i) + \sum_{t = 1}^{\gamma_i} 1(T_i \ge t) \lambda^{se}(t, \gamma_i) + \mu^{de}(\gamma_i)D_i.
\end{equation}

If we further assume Assumption \ref{a1}-(A3), which states that the effect of having one additional treated neighbor remains constant regardless of the number of treated neighbors already present, then \(\lambda^{se}(t, \gamma_i)\) does not depend on \(t\). Thus, we have \(\lambda^{se}(t, \gamma_i) = \lambda^{se}(\gamma_i)\). Consequently, equation \eqref{condE2} further simplifies to:
\[
\mathbf{E}(Y_i \mid D_i, T_i, \gamma_i) = \theta^{00}(\gamma_i) + \mu^{de}(\gamma_i)D_i+\lambda^{se}(\gamma_i) T_i.
\]
Defining $\varepsilon_i\equiv Y_i-\mathbf{E}(Y_i \mid D_i, T_i, \gamma_i)$ completes the proof.

\subsection{Proof for Theorem~\ref{thm2}} \label{proof:thm2}

Under Assumptions \ref{a1} and \ref{a2}, Theorem \ref{thm1} provides the representation that holds true for any $Y_i$:
\[
Y_i = \theta^{00}(\gamma_i) + \mu^{de}(\gamma_i) D_i + \lambda^{se}(\gamma_i) T_i + \varepsilon_i,
\]
while the specified model is:
\begin{equation} \label{sp1}
Y_i = \alpha_0 + \alpha_d D_i + \alpha_t T_i + \alpha_{\gamma} \gamma_i + u_i.
\end{equation}
We first interpret the probability limit of the OLS estimator \(\hat{\alpha}_d\) and then address \(\hat{\alpha}_t\).

\subsection*{Interpretation of \(\alpha_d\)}

For any scalar random variable \(Y\) and \(k \times 1\) random vectors \(W\) with finite second moments, let \(L(Y \mid W)\) be the linear projection of \(Y\) on \(W\), i.e.,
\[
L(Y \mid W) = W' \mathbf{E}(WW')^{-1} \mathbf{E}(WY) = \mathbf{E}(YW') \mathbf{E}(WW')^{-1} W.
\]
Applying \(L(\cdot \mid \gamma_i, T_i)\) to equation \eqref{sp1}, we have:
\[
L(Y_i \mid \gamma_i, T_i) = \alpha_0 + \alpha_d L(D_i \mid \gamma_i, T_i) + \alpha_t T_i + \alpha_{\gamma} \gamma_i.
\]
Subtracting this from equation \eqref{sp1}, we get:
\[
Y_i - L(Y_i \mid \gamma_i, T_i) = \alpha_d (D_i - L(D_i \mid \gamma_i, T_i)) + u_i.
\]
As discussed in \citet{angrist2009}, the OLS estimator \(\hat{\alpha}_d\) has the following probability limit \(\alpha_d\):
\[
\alpha_d = \frac{\text{Cov}(Y_i, D_i - L(D_i \mid \gamma_i, T_i))}{\text{Var}(D_i - L(D_i \mid \gamma_i, T_i))} = \frac{\text{Cov}(Y_i, D_i)}{\text{Var}(D_i)},
\]
where the last equality follows from \(L(D_i \mid \gamma_i, T_i) = \mathbf{E}(D_i) = p\), a constant. Plugging in the true representation of \(Y_i\) from Theorem \ref{thm1}, we have:
\begin{align*}
\alpha_d &= \frac{\text{Cov}(\theta^{00}(\gamma_i), D_i)}{\text{Var}(D_i)} + \frac{\text{Cov}(\mu^{de}(\gamma_i) D_i, D_i)}{\text{Var}(D_i)} + \frac{\text{Cov}(\lambda^{se}(\gamma_i) T_i, D_i)}{\text{Var}(D_i)} \\
&= \frac{\text{Cov}(\mu^{de}(\gamma_i) D_i, D_i)}{\text{Var}(D_i)} \\
&= \frac{\mathbf{E}[\mu^{de}(\gamma_i) D_i (D_i - \mathbf{E}(D_i))]}{\text{Var}(D_i)} \\
&= \frac{\mathbf{E}[\mu^{de}(\gamma_i)]\mathbf{E}[D_i(D_i - \mathbf{E}(D_i))]}{\text{Var}(D_i)} \\
&= \mathbf{E}[\mu^{de}(\gamma_i)],
\end{align*}
where we have used the fact that \(D_i\) is independent of \((\gamma_i, T_i)\).

\subsection*{Interpretation of \(\alpha_t\)}
To interpret the probability limit of the OLS estimator \(\hat{\alpha}_2\), consider the following linear projection:
\[
L(Y_i \mid D_i, \gamma_i) = \alpha_0 + \alpha_d D_i + \alpha_t L(T_i \mid D_i, \gamma_i) + \alpha_{\gamma} \gamma_i,
\]
which gives:
\[
Y_i - L(Y_i \mid D_i, \gamma_i) = \alpha_t (T_i - L(T_i \mid D_i, \gamma_i)) + u_i.
\]
Next, consider the linear projection \(L(T_i \mid D_i, \gamma_i) = \eta_0 + \eta_1 D_i + \eta_2 \gamma_i\) with projection coefficients \(\eta\)'s. Since \(D_i\) and \(\gamma_i\) are uncorrelated, we can apply the linear projection separately to obtain: \(\eta_1 = \frac{\text{Cov}(D_i, T_i)}{\text{Var}(D_i)} = 0\) and \(\eta_2 = \frac{\text{Cov}(T_i, \gamma_i)}{\text{Var}(\gamma_i)} = p\), and \(\eta_0 = \mathbf{E}(T_i) = p \mathbf{E}(\gamma_i)\). Therefore,
\[
L(T_i \mid D_i, \gamma_i) = p \mathbf{E}(\gamma_i) + p \gamma_i.
\]
Define \(\tilde{T}_i \equiv T_i - L(T_i \mid D_i, \gamma_i) = T_i - p \mathbf{E}(\gamma_i) - p \gamma_i = T_i - \mathbf{E}(T_i) - \mathbf{E}(T_i \mid \gamma_i)\). Then,
\begin{align*}
\alpha_t &= \frac{\text{Cov}(Y_i, \tilde{T}_i)}{\text{Var}(\tilde{T}_i)} \\
&= \frac{\text{Cov}(Y_i, T_i - \mathbf{E}(T_i \mid \gamma_i))}{\text{Var}(T_i - \mathbf{E}(T_i \mid \gamma_i))}
\end{align*}
since \(\mathbf{E}(T_i)\) is constant. Plugging in the true model \(Y_i = \theta^{00}(\gamma_i) + \mu^{de}(\gamma_i) D_i + \lambda^{se}(\gamma_i) T_i + \varepsilon_i\), and using the fact that the residual \(T_i - \mathbf{E}(T_i \mid \gamma_i)\) is independent of any function of \(\gamma_i\), and that \(D_i\) is independent of any function of \((\gamma_i, T_i)\), we have:
\begin{align*}
\alpha_t &= \frac{\text{Cov}(\lambda^{se}(\gamma_i) T_i, T_i - \mathbf{E}(T_i \mid \gamma_i))}{\text{Var}(T_i - \mathbf{E}(T_i \mid \gamma_i))} \\
&= \frac{\mathbf{E}[\lambda^{se}(\gamma_i) T_i (T_i - \mathbf{E}(T_i \mid \gamma_i))]}{\mathbf{E}[\{T_i - \mathbf{E}(T_i \mid \gamma_i)\}^2]} \\
&= \frac{\mathbf{E}[\lambda^{se}(\gamma_i) \text{Var}(T_i \mid \gamma_i)]}{\mathbf{E}[\text{Var}(T_i \mid \gamma_i)]},
\end{align*}
where the last equality follows from the law of iterated expectations.

\subsection{Proof for Theorem~\ref{thm3}} \label{proof:thm3}
Recall that the specified model is given by:
\begin{eqnarray}\label{sp2}
Y_i = \eta_0 + \eta_d D_i + \eta_{\bar d} \bar{D}_i^* + u_i.
\end{eqnarray}
The interpretation of \(\eta_d\) is analogous to that of \(\alpha_d\) in the proof of Theorem \ref{thm2}. Therefore, we will focus on the case of \(\eta_{\bar d}\).

\subsection*{Interpretation of \(\eta_{\bar d}\)}
To interpret the probability limit of the OLS estimator \(\hat{\beta}_2\), consider the following linear projection:
\[
L(Y_i \mid D_i) = \eta_0 + \eta_d D_i + \eta_{\bar d} L(\bar{D}_i^* \mid D_i).
\]
After projecting out \(L(Y_i \mid D_i)\) from the equation \ref{sp2}, we have:
\[
Y_i - L(Y_i \mid D_i) = \eta_{\bar d} (\bar{D}_i^* - L(\bar{D}_i^* \mid D_i)) + u_i.
\]
Since \(D_i\) is binary, we have \(L(Y_i \mid D_i) = \mathbf{E}(Y_i \mid D_i)\) and \(L(\bar{D}_i^* \mid D_i) = \mathbf{E}(\bar{D}_i^* \mid D_i)\). Thus,
\[
\eta_{\bar d} = \frac{\text{Cov}(Y_i, \bar{D}_i^* - \mathbf{E}(\bar{D}_i^* \mid D_i))}{\text{Var}(\bar{D}_i^* - \mathbf{E}(\bar{D}_i^* \mid D_i))} = \frac{\text{Cov}(Y_i, \bar{D}_i^* - \mathbf{E}(\bar{D}_i^*))}{\text{Var}(\bar{D}_i^* - \mathbf{E}(\bar{D}_i^*))} = \frac{\text{Cov}(Y_i, \bar{D}_i^*)}{\text{Var}(\bar{D}_i^*)},
\]
where the second equality follows from the fact that \(D_i\) is independent of any function of \((T_i, \gamma_i)\), and thus \(\bar{D}_i^*\) as well.
Plugging in the true \(Y_i\), the numerator of $\eta_{\bar d}$ becomes:
\[
\text{Cov}(\theta^{00}(\gamma_i), \bar{D}_i^*) + \text{Cov}(\mu^{de}(\gamma_i) D_i, \bar{D}_i^*) + \text{Cov}(\lambda^{se}(\gamma_i) T_i, \bar{D}_i^*).
\]
Ideally, the first two terms should be zero, as we are specifically measuring spillover effects. However, this is not the case, as demonstrated below. Recall that:

\[
\mathbf{E}(\bar{D}_i^*) = \mathbf{E}(D_i) \Pr(\gamma_i > 0) \equiv p \cdot p_{\gamma},
\]

then:
\begin{align*}
\text{Cov}(\theta^{00}(\gamma_i), \bar{D}_i^*) &= \mathbf{E}[\theta^{00}(\gamma_i) \bar{D}_i^*] - \mathbf{E}[\theta^{00}(\gamma_i)] \mathbf{E}[\bar{D}_i^*] \\
&= \mathbf{E}[\theta^{00}(\gamma_i) \bar{D}_i^* \mid \gamma_i > 0] p_{\gamma} - \mathbf{E}[\theta^{00}(\gamma_i)] p p_{\gamma} \\
&= \mathbf{E}[\theta^{00}(\gamma_i) \mid \gamma_i > 0] p p_{\gamma} - \mathbf{E}[\theta^{00}(\gamma_i)] p p_{\gamma} \quad (\text{by law of total probability}) \\
&= \left\{\mathbf{E}[\theta^{00}(\gamma_i) \mid \gamma_i > 0] - \mathbf{E}[\theta^{00}(\gamma_i)]\right\} p p_{\gamma} \\
&= \left\{\mathbf{E}[\theta^{00}(\gamma_i) \mid \gamma_i > 0] - \mathbf{E}[\theta^{00}(\gamma_i) \mid \gamma_i = 0]\right\} p (1 - p_{\gamma}) p_{\gamma}.
\end{align*}
\begin{align*}
\text{Cov}(\mu^{de}(\gamma_i) D_i, \bar{D}_i^*) &= \mathbf{E}[\mu^{de}(\gamma_i) D_i \bar{D}_i^*] - \mathbf{E}[\mu^{de}(\gamma_i) D_i] \mathbf{E}[\bar{D}_i^*] \\
&= p \left\{\mathbf{E}[\mu^{de}(\gamma_i) \bar{D}_i^*] - \mathbf{E}[\mu^{de}(\gamma_i)] \mathbf{E}[\bar{D}_i^*]\right\} \\
&= \left\{\mathbf{E}[\mu^{de}(\gamma_i) \mid \gamma_i > 0] - \mathbf{E}[\mu^{de}(\gamma_i)]\right\} p^2 p_{\gamma} \\
&= \left\{\mathbf{E}[\mu^{de}(\gamma_i) \mid \gamma_i > 0] - \mathbf{E}[\mu^{de}(\gamma_i) \mid \gamma_i = 0]\right\} p^2 p_{\gamma} (1 - p_{\gamma}).
\end{align*}
\begin{align*}
\text{Cov}(\lambda^{se}(\gamma_i) T_i, \bar{D}_i^*) &= \text{Cov}(\lambda^{se}(\gamma_i) \gamma_i \bar{D}_i^*, \bar{D}_i^*) \\
&= \mathbf{E}[\lambda^{se}(\gamma_i) \gamma_i \bar{D}_i^* (\bar{D}_i^* - \mathbf{E}(\bar{D}_i^*))].
\end{align*}

While the denominator is:
$
\text{Var}(\bar{D}_i^*) = \mathbf{E}[\bar{D}_i^* (\bar{D}_i^* - \mathbf{E}(\bar{D}_i^*))]$.

Thus, \(\eta_{\bar d}\) is:
\begin{align}
\eta_{\bar d} &= \text{bias} + \frac{\mathbf{E}[\lambda^{se}(\gamma_i) \gamma_i \bar{D}_i^* (\bar{D}_i^* - \mathbf{E}(\bar{D}_i^*))]}{\mathbf{E}[\bar{D}_i^* (\bar{D}_i^* - \mathbf{E}(\bar{D}_i^*))]} \label{eq:beta2}, \\
\text{bias} &= \frac{p p_{\gamma} \left\{\mathbf{E}[\theta^{00}(\gamma_i) \mid \gamma_i > 0] - \mathbf{E}[\theta^{00}(\gamma_i)] + p \left\{\mathbf{E}[\mu^{de}(\gamma_i) \mid \gamma_i > 0] - \mathbf{E}[\mu^{de}(\gamma_i)]\right\}\right\}}{\text{Var}(\bar{D}_i^*)} \notag,
\end{align}
where \(\text{Var}(\bar{D}_i^*) = p p_{\gamma} \left\{p (1 - p_{\gamma}) + (1 - p) \mathbf{E}\left(\frac{1}{\gamma_i} \mid \gamma_i > 0\right)\right\}\). Therefore,
\begin{align}\label{eq:bias}
\text{bias} &= \frac{(\Delta \theta^{00} + p \Delta \mu^{de})(1 - p_{\gamma})}{p (1 - p_{\gamma}) + (1 - p) \mathbf{E}\left(\frac{1}{\gamma_i} \mid \gamma_i > 0\right)},
\end{align}
where
\begin{eqnarray*}
\Delta \theta^{00} \equiv \mathbf{E}[\theta^{00}(\gamma_i) \mid \gamma_i > 0] - \mathbf{E}[\theta^{00}(\gamma_i) \mid \gamma_i = 0], \quad \Delta \mu^{de} \equiv \mathbf{E}[\mu^{de}(\gamma_i) \mid \gamma_i > 0] - \mathbf{E}[\mu^{de}(\gamma_i) \mid \gamma_i = 0],
\end{eqnarray*}
as claimed.

\subsection{Proof for Theorem \ref{lemma1}: Zero Bias without Isolated Nodes}\label{proof:lemma1}
Suppose there are no isolated nodes, so that \(\bar{D}_i^* = \bar{D}_i\) for all \(i\). In this case, the bias term in equation \eqref{eq:bias} becomes zero as \(p_{\gamma}=0\), whereas the denominator of the second term in equation \eqref{eq:beta2} becomes:
\[
\text{Var}(\bar{D}_i) = \mathbf{E}(\text{Var}(\bar{D}_i \mid \gamma_i)) + \text{Var}(\mathbf{E}(\bar{D}_i \mid \gamma_i)) = \mathbf{E}(\text{Var}(\bar{D}_i \mid \gamma_i)) = \mathbf{E}\left(\frac{p(1 - p)}{\gamma_i}\right).
\]
Therefore, equation \eqref{eq:beta2} becomes:
\[
\eta_{\bar d} = \frac{\mathbf{E}[\lambda^{se}(\gamma_i) \gamma_i \text{Var}(\bar{D}_i \mid \gamma_i)]}{\mathbf{E}(\text{Var}(\bar{D}_i \mid \gamma_i))} = \frac{\mathbf{E}[\lambda^{se}(\gamma_i) \gamma_i \frac{1}{\gamma_i}]}{\mathbf{E}\left(\frac{1}{\gamma_i}\right)} = \mathbf{E}[w_{\bar d}(\gamma_i) \gamma_i\lambda^{se}(\gamma_i)],
\]
where \(w_{\bar d}(\gamma_i) = \frac{\frac{1}{\gamma_i}}{\mathbf{E}\left(\frac{1}{\gamma_i}\right)}\), as claimed.


\end{appendix}

\end{document}